\documentstyle[12pt]{article}

\newcommand{\nn}{\nonumber}
\newcommand{\be}{\begin{equation}}
\newcommand{\ee}{\end{equation}}
\newcommand{\bea}{\begin{eqnarray}}
\newcommand{\eea}{\end{eqnarray}}
\newcommand{\bean}{\begin{eqnarray*}}
\newcommand{\eean}{\end{eqnarray*}}

\begin{document}

\vspace{-1cm}
\noindent
\begin{flushright}
KANAZAWA-03-09\\
KUNS-1827
\end{flushright}
\vspace{10mm}
\begin{center}
{\Large \bf
Exact ${\cal S}_3$ symmetry solving the supersymmetric flavor problem
}
\vspace*{15mm}\\
\renewcommand{\thefootnote}{\alph{footnote}}
Tatsuo Kobayashi$^\dagger$,
\footnote{E-mail: kobayash@gauge.scphys.kyoto-u.ac.jp}
Jisuke Kubo$^{\dagger\dagger}$
\footnote{E-mail: jik@hep.s.kanazawa-u.ac.jp}
and Haruhiko Terao$^{\dagger\dagger}$
\footnote{E-mail: terao@hep.s.kanazawa-u.ac.jp}
\vspace*{5mm}\\
$^\dagger$Department of Physics, Kyoto University\\
Kyoto 606-8502, Japan
\vspace{2mm}\\
$^{\dagger\dagger}$Institute for Theoretical Physics, Kanazawa 
University\\
Kanazawa 920-1192, Japan
\end{center}
\vspace*{10mm}
\begin{abstract}
The exact discrete flavor symmetry,
$S_3$ in the hadronic sector and $S_3 \times Z_2$
in the leptonic sector, which has been
recently found, is introduced in a
supersymmetric extension of
the standard model. We investigate
the supersymmetric flavor problem, and
explicitly  find that thanks to
the flavor symmetry the dangerous
FCNC processes and CP-violating phases are
sufficiently suppressed.
\end{abstract}
\vspace*{20mm}
\noindent
PACS numbers: 12.60.J,11.30.Hv,12.15.F,14.60.Pq

\newpage
\pagestyle{plain}
\pagenumbering{arabic}
\setcounter{footnote}{0}

The minimal supersymmetric standard model (MSSM) \cite{susy}
contains more than 100 free parameters \cite{dimopoulos1}.
Most of them belong to the supersymmetry
breaking (SSB) sector.
The problem is not only this huge number,
but also the fact that
these parameters have to be highly fine tuned
   so that they do not cause
problems with experimental observations
on the flavor changing neutral current
(FCNC) processes and CP-violation
phenomena \cite{fcnc-mueg}-\cite{fcnc}.
This problem,
called the SUSY flavor problem, is not new, but has existed ever,
since supersymmetry found phenomenological
applications.
There are several approaches to overcome this
problem \cite{gauge}-\cite{ babu1}.

The possibility of introducing
more than 100 independent
parameters into the MSSM is closely related
to the  fact that the degrees of freedom inherent in
the Yukawa sector of the standard model (SM)
are much more than the observable degrees of freedom.
Since an exact flavor symmetry in the Yukawa sector
can reduce this redundancy of the SM,
the same symmetry can
reduce that huge number of the
independent parameters of the MSSM,
and could suppress the dangerous FCNC processes and CP violating
phases.
Recently, it has been found \cite{kubo} that
a nonabelian,
discrete flavor symmetry based on the permutation
group $S_3$ is consistent with
the present experimental knowledge.
That is, it is  spontaneously broken,
only because the electroweak gauge symmetry
$SU(2)_L\times U(1)_Y$ is spontaneously broken
\footnote{The permutation symmetries have been considered
   in \cite{hall}, \cite{pakvasa1}-\cite{ma}.}.
The idea to use nonabelian discrete symmetries to
soften the SUSY flavor problem has been in fact proposed in
\cite{hall,hamaguchi}.
However,  the symmetries considered in \cite{hall,hamaguchi}
are explicitly broken at the supersymmetry breaking scale.
Therefore, there is no convincing reason why they should be intact
in the SSB sector.
In this letter we will demonstrate  that the exact flavor symmetry
of \cite{kubo},
$S_3$ in the hadronic sector and $S_3\times Z_2$
in the leptonic sector, can considerably soften the SUSY flavor problem.

Let us begin by supersymmetrizing the $S_3$ invariant SM.
The three generations of quark and lepton chiral superfields
in MSSM are assumed to belong in a reducible representation
of $S_3$ equivalent to \mbox{\boldmath{$1_S +2$ }}.
We also introduce the $S_3$-doublet Higgs pair,
$H^U_I, H^D_I (I=1, 2)$, as well as the
$S_3$-singlet Higgs pair, $H^U_3, H^D_3$.
The same R-parity is assigned to these fields as in MSSM.
Then we assume that the superpotential in MSSM,
$W = W_D + W_U + W_E + W_{\nu}$, is
invariant under the $S_3$-symmetry.
The each part is given explicitly as follows
(the neutrino superpotential is abbreviated for the
present purpose);
\bea
W_D &=&
Y_1^D Q_I H^{D}_3 D_{IR} + Y_3^D Q_3 H^{D}_3 D_{3R}  \nn \\
& &+ Y^{D}_{2}\left[ Q_{I} \kappa_{IJ} H^{D}_1  D_{JR}
+ Q_{I} \eta_{IJ} H^{D}_2  D_{JR} \right] \nn\\
& & + Y^D_{4} Q_3 H^D_I  D_{IR}
+ Y^D_{5} Q_I H^D_I D_{3R},
\label{potd}\\
W_U &=&
Y_1^U Q_I H^{U}_3 U_{IR} + Y_3^U Q_3 H^{U}_3 U_{3R}  \nn \\
& &+ Y^{U}_{2}\left[ Q_{I} \kappa_{IJ} H^{U}_1  U_{JR}
+ Q_{I} \eta_{IJ} H^{U}_2  U_{JR} \right] \nn\\
& & + Y^U_{4} Q_3 H^U_I  U_{IR}
+ Y^U_{5} Q_I H^U_I U_{3R},
\label{potu}\\
W_E &=&
Y_1^E L_I H^{D}_3 E_{IR} + Y_3^E L_3 H^{D}_3 E_{3R}  \nn\\
& &+Y^{E}_{2}\left[ L_{I} \kappa_{IJ} H^{D}_1  E_{JR}
+ L_{I} \eta_{IJ} H^{D}_2  E_{JR} \right] \nn\\
& & + Y^E_{4} L_3 H^D_I  E_{IR}
+ Y^E_{5} L_I H^D_I E_{3R},
\label{pote}
\eea
where
\be
\kappa = \left( \begin{array}{cc}
0& 1\\ 1 & 0\\
\end{array}\right) ~~\mbox{and}~~
\eta = \left( \begin{array}{cc}
1& 0\\ 0 & -1\\
\end{array}\right).
\label{kappa}
\ee
We will discuss the Higgs part of the superpotential later on.
It will be
shown  that Higgs VEV's can satisfy
$\langle H_1 \rangle = \langle H_2 \rangle$,
which we shall assume throughout this letter.
Consequently,  the quark and lepton mass matrices
take the general form
\be
{\bf M}_a = \left( \begin{array}{ccc}
m^a_1+m^a_{2} & m^a_{2} & m^a_{5}
\\  m^a_{2} & m^a_1-m^a_{2} &m^a_{5}
    \\ m^a_{4} & m^a_{4}&  m^a_3
\end{array}\right),
\label{general-m}
\ee
where $a=u, d, e$.
It has been found in \cite{kubo} that the exact
$S_3$ flavor symmetry (along with an additional exact $Z_2$
in the leptonic sector)  is consistent with all the observed
quark and lepton masses and mixing angles.
Since this result can remain valid after supersymmetrization of the
model,
   we would like to focus on  the flavor structure in the
superpartner sector, assuming the same flavor symmetry
in the SSB sector.

Now we come to discuss the SUSY flavor problem in the slepton sector.
The $Z_2$-parity assignment is the same  as in \cite{kubo}:
\bea
+&  & \mbox{ for} ~~H_3^{U,D}, ~L_3, ~L_I, ~E_{3R},~ E_{IR}~
~\mbox{ and}~~
- ~~\mbox{for}~~H_I^{U,D}.
\eea
Note that $m_1^e=m_3^e = 0$ due to this discrete symmetry.
Therefore,  the lepton mass
matrix can be written as
\be
{\bf M}_e = m^e_2 \left(
\begin{array}{ccc}
1 & 1 & x \\
1 & -1 & x \\
y & y & 0
\end{array}
\right),
\label{leptonmass}
\ee
where $x = m_5^e/m_2^e = 16.78$ and
$y = m_4^e/m_2^e = 0.006836$ \cite{kubo}.

In order to evaluate the amount of the flavor changing
processes generated in this model, we need to know the
unitary matrices $U_{eL}$ and $U_{eR}$
to the mass eigenstates, which satisfy
\be
U_{eL}^{\dagger} {\bf M}_e U_{eR}
= \mbox{diag}( m_e, m_{\mu}, m_{\tau}).
\ee
It is noticed that $x \simeq m_{\tau}/m_{\mu}$ and
$y \simeq \sqrt{2}m_e/m_{\mu}$.
Since $y$ and $1/x$ are small numbers, let us
evaluate the rotation matrices by
expanding them with respect to these parameters.
Here we evaluate the matrices at $O(y^2/x^2)$.
Then the rotation matrices are found to be
\bea
U_{eL} &= &\left(
\begin{array}{ccc}
\frac{y}{2}\left( 1 + \frac{1}{x^2}\right) &
\frac{1}{\sqrt{2}} \left(1-\frac{y^2}{4}-\frac{y^2}{2x^2} \right) &
\frac{1}{\sqrt{2}} \\
-\frac{y}{2}\left( 1 - \frac{1}{x^2}\right) &
-\frac{1}{\sqrt{2}} \left(1-\frac{y^2}{4}+\frac{y^2}{2x^2} \right) &
\frac{1}{\sqrt{2}} \\
-1+\frac{y^2}{4} & \frac{y}{\sqrt{2}} &  \frac{y}{\sqrt{2}x^2}
\end{array}\right),
\label{UeL}
\\
U_{eR} &= &\left(
\begin{array}{ccc}
-1 +\frac{1}{2x^2} & \frac{y^2}{2}\left( 1 - \frac{1}{x^2}\right) &
\frac{1}{x} \\
\frac{y^2}{2}\left( 1 - \frac{1}{2 x^2}\right) & 1  & 0 \\
\frac{1}{x} & -\frac{y^2}{2x} &  1 - \frac{1}{2x^2}
\end{array}\right).
\label{UeR}
\eea
The MNS matrix 
$V_{\rm MNS} \equiv U_{eL}^{\dagger}U_{\nu}$ is also
found by using the rotation matrix of the neutrino
sector $U_{\nu}$ given in ref.~\cite{kubo}.
Note that the maximal mixing appearing in the left handed
rotation matrix is responsible for the atmospheric neutrino
mixing. 
While the solar neutrino mixing is explained by the
large mixing in the rotation matrix of the neutrino mass matrix.
That is, the elements,  $(U_{eL})_{21}$ and $(U_{eL})_{23}$
become, respectively, the $(1,3)$
and $(3,3)$ elements of the mixing matrix $V_{\rm MNS}$;
\bea
U_{e3} &=&-y(1-1/x^2)/2 \simeq -m_e/\sqrt{2}m_\mu~,~
\cos\theta_{\rm atm}=1/\sqrt{2}.
\eea

The $S_3$ symmetry restricts also the soft scalar mass matrices
so that the off-diagonal elements are forbidden by the
symmetry and two of three are completely degenerate.
Explicitly the matrices are given in the form of
\be
{\bf \tilde{m}^2}_{LL} =
m^2_{\tilde{\ell}} \left(
\begin{array}{ccc}
a_L & 0 & 0 \\
0 & a_L & 0 \\
0 & 0 & b_L
\end{array}
\right),~~~~~~
{\bf \tilde{m}^2}_{RR} =
m^2_{\tilde{\ell}} \left(
\begin{array}{ccc}
a_R & 0 & 0 \\
0 & a_R & 0 \\
0 & 0 & b_R
\end{array}
\right),
\label{scalarmass}
\ee
where $m_{\tilde{\ell}}$ denotes the average of the
slepton masses and $(a_{L(R)}, b_{L(R)})$ are
free parameters.

The branching ratio of a LFV event, {\it e.g.}
$Br(\mu \rightarrow e + \gamma)$,
is proportional to the off-diagonal elements of the mass matrix
with respect to the lepton mass eigenstates, which
is given by
$U_{eL}^{\dagger} {\bf \tilde{m}^2}_{LL} U_{eL}$.
By using the expression given is by eq.~(\ref{UeL}),
this matrix may be evaluated as
\be
U_{eL}^{\dagger}{\bf \tilde{m}^2}_{LL} U_{eL}
  \simeq  m^2_{\tilde{\ell}} \left(
\begin{array}{ccc}
b_L - (y^2/2) \Delta_L
& -(y/\sqrt{2}) \Delta_L
& -(y/\sqrt{2}x^2) \Delta_L \\
-(y/\sqrt{2}) \Delta_L
& a_L + (y^2/2) \Delta_L
& (y^2/2x^2) \Delta_L \\
-(y/\sqrt{2}x^2) \Delta_L
& (y^2/2x^2) \Delta_L
&  a_L
\end{array}\right),
\label{m2LL}
\ee
where $\Delta_L = b_L - a_L$.
It is convenient to see the ratio of the off-diagonal
term to average of the diagonal terms, 
$(\delta^{\ell}_{ij})_{LL}$,
whose upper bounds from experimental results are 
given in the literatures \cite{fcnc}.
The explicit numbers of $\delta^{\ell}$'s in this model
are predicted as follows;
\bea
(\delta^{\ell}_{12})_{LL}
&\simeq&  -4.8 \times 10^{-3} ~\Delta_L, \\
(\delta^{\ell}_{13})_{LL}
&\simeq&  -1.7 \times 10^{-5} ~\Delta_L, \\
(\delta^{\ell}_{23})_{LL}
&\simeq&  8.3 \times 10^{-8} ~\Delta_L.
\eea

The upper bounds for ratios of the off-diagonal elements
$(\delta_{ij})_{LL(RR)}$ \cite{fcnc} are shown in Table~1.
It is seen that all the off-diagonal elements
are small enough to satisfy the LFV constraints.
It is noted that the $(1,2)$- and
$(1,3)$-elements are suppressed due to the small mixing.
On the other hand the slepton mass matrix for the $S_3$-doublets,
which are roughly regarded as smu and stau,
is proportional to the identity matrix.
Therefore the $(2,3)$-element is almost vanishing,
although the mixing angle of these bases is maximum.

\begin{table}[thb]
\begin{center}
\begin{tabular}{|c|c|} \hline \hline
$|(\delta^{\ell}_{12})_{LL}|$ & $|(\delta^{\ell}_{12})_{LR}|$
\\ \hline
$ 7.7 \times 10^{-3} ~\tilde{m}^2_{\tilde{\ell}} $
& $ 1.7 \times 10^{-6} ~\tilde{m}^2_{\tilde{\ell}} $
\\ \hline \hline
$|(\delta^{\ell}_{13})_{LL}|$ & $|(\delta^{\ell}_{13})_{LR}|$
\\ \hline
$ 29 ~ \tilde{m}^2_{\tilde{\ell}} $
& $ 1.1 \times 10^{-1} ~ \tilde{m}^2_{\tilde{\ell}} $
\\ \hline \hline
$|(\delta^{\ell}_{23})_{LL}|$ & $|(\delta^{\ell}_{23})_{LR}|$
\\ \hline
$ 5.3  ~ \tilde{m}^2_{\tilde{\ell}} $
& $ 2.0 \times 10^{-2}  ~ \tilde{m}^2_{\tilde{\ell}} $
\\ \hline \hline
\end{tabular}
\caption{Limits on the $|\delta^{\ell}_{ij}|$ from
$\ell_i \rightarrow \ell_i \gamma$ decays for
$m^2_{\tilde{\gamma}}/m^2_{\tilde{\ell}} = 1$.
Here the parameter $\tilde{m}_{\tilde{\ell}}$ denotes
$m_{\tilde{\ell}}(\mbox{GeV})/100$.
See \cite{fcnc} for details.}
\end{center}
\end{table}

Similarly, for the right-handed slepton the off-diagonal
elements of the mass matrix may be evaluated by using
$U_{eR}$ given by eq.~(\ref{UeR}) and found to be
\be
U_{eR}^{\dagger}{\bf \tilde{m}^2}_{RR} U_{eR}
\simeq m^2_{\tilde{\ell}}\left(
\begin{array}{ccc}
a_R + (1/x^2) \Delta_R & -(y^2/2x^2) \Delta_R &
(1/x) \Delta_R \\
-(y^2/2x^2) \Delta_R & a_R &  -(y^2/2x) \Delta_R \\
(1/x)\Delta_R & -(y^2/2x) \Delta_R
&  b_R - (1/x^2) \Delta_R
\end{array}\right),
\label{m2RR}
\ee
where $\Delta_R = b_R - a_R$
The explicit ratios of the off-diagonal elements to the
diagonal elements are found to be
\bea
(\delta^{\ell}_{12})_{RR}
&\simeq& -8.28 \times 10^{-8} ~\Delta_R, \\
(\delta^{\ell}_{13})_{RR}
&\simeq& 5.49 \times 10^{-2} ~\Delta_R, \\
(\delta^{\ell}_{23})_{RR}
&\simeq& - 1.39 \times 10^{-6} ~\Delta_R.
\eea
Thus the flavor mixing in the right-handed sector is also
very small.

It is also necessary for the mass matrix between the
left-handed and the right-handed sleptons, which
is generated through the  so-called A-terms,
to satisfy the FCNC constraints.
Since the Yukawa interactions in the superpotential
must be invariant under the $S_3$-symmetry,
the left-right mass matrix may be parametrized as
\be
{\bf \tilde{m}^2}_{LR} =  \left(
\begin{array}{ccc}
m_2 A_2 & m_2 A_2 & m_5 A_5 \\
m_2 A_2 & -m_2 A_2 & m_5 A_5 \\
m_4 A_4 & m_4 A_4 & 0
\end{array}
\right),
\ee
where $A_i$ are free parameters. Here, however, we also
assume them to be on the order of the gaugino masses.
After rotating to the bases of lepton mass eigenstates,
the mass matrix is found to be
\bea
\lefteqn{
U_{eL}^{\dagger}{\bf \tilde{m}^2}_{LR}U_{eR}
} \nn \\
&\simeq&  m_2 \left(
\begin{array}{ccc}
y A_4 & y (A_2 - A_4) & -(y/x)(A_4 - A_5) \\
-(y^2/\sqrt{2})(A_2 - A_4)
& \sqrt{2}A_2 & (y^2/\sqrt{2}x) (A_4-A_5) \\
-\sqrt{2}(A_2 - A_5) & (y^2/\sqrt{2}) (A_2-A_5)
& \sqrt{2}x A_5
\end{array}
\right)
\eea
The ratio of the off-diagonal term to the average
slepton mass $m_{\tilde{\ell}}$ are explicitly
given by
\bea
(\delta^{\ell}_{12})_{LR}
&\simeq& 5.1 \times 10^{-6} ~
(\tilde{A}_2-\tilde{A}_4)
\left( \frac{100}{m_{\tilde{\ell}}(\mbox{GeV})}\right)^2, \\
(\delta^{\ell}_{13})_{LR}
&\simeq& 1.1 \times 10^{-2} ~
(\tilde{A}_2-\tilde{A}_5)
\left( \frac{100}{m_{\tilde{\ell}}(\mbox{GeV})}\right)^2, \\
(\delta^{\ell}_{23})_{LR}
&\simeq& 2.5 \times 10^{-8} ~
(\tilde{A}_2-\tilde{A}_5)
\left( \frac{100}{m_{\tilde{\ell}}(\mbox{GeV})}\right)^2,
\eea
where $\tilde{A}_i$ denotes $A_i(\mbox{GeV})/100$.
These results should be compared with the limits
for $|\delta^{\ell}_{LR}|$ shown in Table~1.
It is seen that $(\delta^{\ell}_{12})_{LR}$ saturates
nearly the experimental bound from
$\mu \rightarrow e~ \gamma$ process,
unless $A_2-A_4$ is somewhat suppressed with some reason.
The other off-diagonal elements are sufficiently smaller
than the experimental bounds.
Indeed alignment of the A-terms to the Yukawa couplings
is fairly good in this model.
The origin of this remarkable feature is that
the structure of A-terms are also restricted by the
$S_3$-symmetry.

The additional discrete symmetry
$Z_2$ (which turns out be  is anomaly-free)  forbids CP-violations in 
the leptonic
as well as in the hadronic sector.
Consequently, $Z_2$ should be explicitly broken in the hadronic sector
to accommodate CP violations.
Therefore, the mass matrices for the up (down) sector quarks are
parametrized as given by eq.~(\ref{general-m}).
As in  the leptonic case, we introduce the unitary
matrices $U_{u(d)L}$ and  $U_{u(d)R}$ satisfying
\be
U_{u(d)L}^{\dagger} {\bf M}_{u(d)} U_{u(d)R} =
\mbox{diag} (m_{u(d)}, m_{c(s)}, m_{t(b)}).
\ee
In \cite{kubo}, it has been found that the
hierarchical quark masses in the up and down sectors,
\bea
m_u/m_t &=&1.33\times 10^{-5}~,~m_c/m_t=2.99\times 10^{-3},\nn\\
m_d/m_b &=&1.31\times 10^{-3}~,~m_s/m_b=1.17\times 10^{-2},
\eea
as well as the parameters in the
CKM matrix may be obtained by choosing the parameters
in the mass matrices as follows;
\bea
m_1^u/m_3^u &=& -0.000293~,~ m_2^u/m_3^u =-0.00028,\nn\\
m_4^u/m_3^u &=&0.031  ~,~
m_5^u/m_3^u=0.0386,\nn \\
m_1^d/\mbox{Re}(m_3^d) &=&0.0004~,~  
m_2^d/\mbox{Re}(m_3^d) =0.00275~, \nn \\
m_3^d/\mbox{Re}(m_3^d) &=& 1+ 1.2I~,~
m_4^d/\mbox{Re}(m_3^d)  = 0.283~,  \nn \\
m_5^d /\mbox{Re}(m_3^d) &=& 0.058.
\label{choice1}
\eea
It should be noted that there are $10$ real parameters
and one phase in mass matrices to fine tune
to produce  six quark masses, three mixing angles and one
CP-violating phase. Therefore the unitary matrices may be
determined without ambiguity in this tuning process.
The explicit unitary matrices are found to be
\bea
& &
U_{uL} \simeq\left(
\begin{array}{ccc}
0.64 & -0.77 & 0.038 \\
-0.77 & -0.64 & 0.038 \\
0.0051 & 0.054 & 1.0
\end{array}\right), 
\label{UuL} \\
& &
U_{uR} \simeq \left(
\begin{array}{ccc}
0.64 & 0.77 & 0.031 \\
-0.77 & 0.64 & 0.031 \\
0.0041 & -0.043 & 1.0
\end{array}\right), 
\label{UuR} \\
& &
U_{dL}  \simeq \left(
\begin{array}{ccc}
0.77 & 0.62 - 0.15~I & 0.023 -0.027~I \\
-0.62 - 0.15~I & 0.77 & 0.022 - 0.027~I \\
-0.0079- 0.00063~I  & -0.036-0.034~I & 1.0
\end{array}\right), 
\label{UdL} \\
& &
U_{dR}  \simeq \left(
\begin{array}{ccc}
0.77 & 0.59+0.16~I & 0.11 + 0.13~I \\
-0.62 + 0.15~I & 0.75 &  0.11 + 0.13~I \\
-0.039 + 0.0031~I & -0.18 + 0.17~I & 1.0
\end{array}\right),
\label{UdR}
\eea
Note that the off-diagonal elements in $U_{dL}$ and $U_{dR}$ carry
large complex phases.

The $S_3$-symmetry restricts also the squark mass matrices
to the same forms as given by eq.~(\ref{scalarmass}).
These matrices should be rotated to bases of the quark mass
eigenstates by using the unitary matrices given by
eqs.~(\ref{UuL}-\ref{UdR}).
As for the down-type squark mass matrix,
the ratios of the off-diagonal elements to the average of
squark mass squared turn out to be as follows;
\bea
(\delta^d_{12})_{LL}
& \simeq&  (3.0 \times 10^{-4}+2.4 \times 10^{-4}~I) ~\Delta_L, \\
(\delta^d_{13})_{LL}
& \simeq&  (-7.9\times 10^{-3}+6.3 \times 10^{-4}~I) ~\Delta_L, \\
(\delta^d_{23})_{LL}
& \simeq&  (-3.5 \times 10^{-2}+ 3.4 \times 10^{-2}~I) ~\Delta_L.
\eea
and
\bea
(\delta^d_{12})_{RR}
& \simeq& (7.3 \times 10^{-3} - 6.0 \times 10^{-3}~I) ~\Delta_R, \\
(\delta^d_{13})_{RR}
& \simeq& (- 3.7 \times 10^{-2} - 3.0 \times 10^{-3}~I) ~\Delta_R, \\
(\delta^d_{23})_{RR}
& \simeq& (-1.7 \times 10^{-1} -1.6 \times 10^{-1}~I) ~\Delta_R.
\eea
where $\Delta_{L(R)} = b_{L(R)} - a_{L(R)}$.
The upper bounds for the parameters from measurements of
$K-\bar{K}$, $D-\bar{D}$, $B_d-\bar{B}_d$  mixing,
$\epsilon_K$, $b \rightarrow s~ \gamma$ and
$\epsilon'/\epsilon$ \cite{fcnc}
are shown in Table~2.
The imaginary parts are constrained by CP-violating processes.
Interestingly enough these mixings satisfy the experimental
bounds not only for the FCNC but also for CP-violation.
Also the off diagonal elements of the up-type squark mass 
matrix are found to be on the same order of
those for the down-type squarks. 
Therefore mixing in the up-type squark sector is sufficiently 
smaller than the experimental bounds.

\begin{table}[htb]
\begin{center}
\begin{tabular}{|c|c|c|} \hline \hline
$\sqrt{|\mbox{Re}(\delta^d_{12})^2_{LL,RR}|}$
& $\sqrt{|\mbox{Re}(\delta^d_{12})_{LL}(\delta^d_{12})_{RR}|}$
& $\sqrt{|\mbox{Re}(\delta^d_{12})^2_{LR}|}$
\\ \hline
$4.0 \times 10^{-2} ~\tilde{m}_{\tilde{q}}$
& $2.8 \times 10^{-3} ~\tilde{m}_{\tilde{q}}$
& $4.4 \times 10^{-3} ~\tilde{m}_{\tilde{q}}$
\\ \hline \hline
$\sqrt{|\mbox{Re}(\delta^d_{13})^2_{LL,RR}|}$
& $\sqrt{|\mbox{Re}(\delta^d_{13})_{LL}(\delta^d_{13})_{RR}|}$
& $\sqrt{|\mbox{Re}(\delta^d_{13})^2_{LR}|}$
\\ \hline
$9.8 \times 10^{-2} ~\tilde{m}_{\tilde{q}}$
& $1.8 \times 10^{-2} ~\tilde{m}_{\tilde{q}}$
& $3.3 \times 10^{-3} ~\tilde{m}_{\tilde{q}}$
\\ \hline \hline
$\sqrt{|\mbox{Re}(\delta^u_{12})^2_{LL,RR}|}$
& $\sqrt{|\mbox{Re}(\delta^u_{12})_{LL}(\delta^u_{12})_{RR}|}$
& $\sqrt{|\mbox{Re}(\delta^u_{12})^2_{LR}|}$
\\ \hline
$1.0 \times 10^{-1} ~\tilde{m}_{\tilde{q}}$
& $1.7 \times 10^{-2} ~\tilde{m}_{\tilde{q}}$
& $3.1 \times 10^{-3} ~\tilde{m}_{\tilde{q}}$
\\ \hline \hline
$\sqrt{|\mbox{Im}(\delta^d_{12})^2_{LL,RR}|}$
& $\sqrt{|\mbox{Im}(\delta^d_{12})_{LL}(\delta^d_{12})_{RR}|}$
& $\sqrt{|\mbox{Im}(\delta^d_{12})^2_{LR}|}$
\\ \hline
$3.2 \times 10^{-3}~ \tilde{m}_{\tilde{q}}$
& $2.2 \times 10^{-4} ~\tilde{m}_{\tilde{q}}$
& $3.5 \times 10^{-4} ~\tilde{m}_{\tilde{q}}$
\\ \hline \hline
$|(\delta^d_{23})_{LL,RR}|$
&
& $|(\delta^d_{23})_{LR}|$
\\ \hline
$8.2~ \tilde{m}_{\tilde{q}}^2$
&
& $1.6 \times 10^{-2} ~\tilde{m}_{\tilde{q}}^2$
\\ \hline \hline
$|\mbox{Im}(\delta^d_{12})_{LL,RR}|$
&
& $|\mbox{Im}(\delta^d_{12})_{LR}|$
\\ \hline
$4.8 \times 10^{-1} ~\tilde{m}_{\tilde{q}}^2$
&
& $2.0 \times 10^{-5} ~\tilde{m}_{\tilde{q}}^2$
\\ \hline \hline
\end{tabular}
\caption{Limits on the $|\delta^{d(u)}_{ij}|$ from
$K-\bar{K}$, $D-\bar{D}$, $B_d-\bar{B}_d$  mixing,
$\epsilon_K$, $b \rightarrow s~ \gamma$ and
$\epsilon'/\epsilon$ for
$m_{\tilde{g}}/m_{\tilde{q}}=1$ \cite{fcnc}.
Here the parameter $\tilde{m}_{\tilde{q}}$ denotes
$m_{\tilde{q}}(\mbox{GeV})/500$.}
\end{center}
\end{table}

Mixing between the left-handed and the right-handed squarks
and also their effects to FCNC and CP-violation may be
evaluated just as done for the slepton sector.
Again it is crucial that the A-terms are constrained
by the $S_3$-symmetry. The left-right mass matrix
is parametrized as
\be
{\bf \tilde{m}^2}_{LR} =  \left(
\begin{array}{ccc}
m_1 A_1 + m_2 A_2 & m_2 A_2 & m_5 A_5 \\
m_2 A_2 & m_1 A_1 - m_2 A_2 & m_5 A_5 \\
m_4 A_4 & m_4 A_4 & m_3 A_3
\label{qLRmass}
\end{array}
\right),
\ee
with assuming $A_i$ are on the same order of the gaugino mass.
Now the off-diagonal elements on the bases of quark mass
eigenstates may be explicitly obtained, since the
rotation matrices $U_{uL(R)}$ and $U_{dL(R)}$ are fixed.
Their ratio to the average squark masses, whose experimental
constraints are shown also in Table~2, are found to be
\bea
(\delta^{u}_{12})_{LR}
&\sim& O(10^{-5})
\left( \frac{500}{m_{\tilde{q}}(\mbox{GeV})}\right)^2, \\
(\delta^{u}_{13})_{LR}
&\sim& O(10^{-4})
\left( \frac{500}{m_{\tilde{q}}(\mbox{GeV})}\right)^2, \\
(\delta^{u}_{23})_{LR}
&\sim& O(10^{-3})
\left( \frac{500}{m_{\tilde{q}}(\mbox{GeV})}\right)^2, \\
(\delta^{d}_{12})_{LR}
&\sim& \left(O(10^{-6}) + O(10^{-8})~I \right)
\left( \frac{500}{m_{\tilde{q}}(\mbox{GeV})}\right)^2, \\
(\delta^{d}_{13})_{LR}
&\sim& \left(O(10^{-5}) + O(10^{-5})~I \right)
\left( \frac{500}{m_{\tilde{q}}(\mbox{GeV})}\right)^2, \\
(\delta^{d}_{23})_{LR}
&\sim& \left(O(10^{-4}) + O(10^{-5})~I \right)
\left( \frac{500}{m_{\tilde{q}}(\mbox{GeV})}\right)^2,
\eea
where the A-parameters $A_i$ are assumed to be $O(500\mbox{GeV})$.
It is seen that alignment of the LR-mass matrix to the Yukawa
matrix is remarkably good. 
This is because (1,2)-elements of the matrix
${\bf \tilde{m}^2}_{LR} ({\bf \tilde{m}^2}_{LR})^{\dagger}$,
or $({\bf \tilde{m}^2}_{LR})^{\dagger} {\bf \tilde{m}^2}_{LR}$,
are almost of the democratic type and mixing
with the 3rd elements are tiny.
Thus it has been found that
the FCNC as well as CP constraints for supersymmetric
extension of the standard model are resolved solely by
the exact flavor symmetry of $S_3$.

One of the important features of this model is to have a
pair of $S_3$-doublet Higgs fields $H_I^{U(D)} (I=1, 2)$
as well as a pair of $S_3$-singlet Higgs $H_3^{U(D)}$.
Also all the Higgs fields must acquire non-vanishing
VEV. Moreover $\langle H_1^{U(D)} \rangle =
\langle H_2^{U(D)} \rangle$ should be satisfied, which
has been assumed so far.
On the other hand the $S_3$-invariant supersymmetric
mass terms for these
Higgs fields may be given by
\be
W = \mu_D H_I^U H_I^D + \mu_S H_3^U H_3^D,
\label{mu-term}
\ee
as extension of the so-called $\mu$-term in the MSSM.
Then, however, it is found that the global symmetry of the tree-level
Higgs potential
 is enhanced to the complex extension of $U(2)$.
Consequently,  there appear extra massless goldstone particles
and this naive extension cannot be accepted
phenomenologically.
Note, however, that  the whole theory does not have this enhanced symmetry
and their masses are not protected from radiative
corrections in the presence of  soft supersymmetry breaking terms.
Therefore,  the dangerous light Higgs particles are not exactly
massless (pseudo-goldstones), although
they will be  still much lighter than the
lightest Higgs in the MSSM.

Thus it is necessary to extend the Higgs sector so that the tree
level potential does not have continuous symmetries enhanced from
$S_3$. For this purpose let us introduce SM-gauge singlets
$N_I (I=1,2)$ and $N_3$ belonging to
\mbox{\boldmath{$2 + 1_S $ }} of $S_3$, where we 
assume that $N$'s have even parity of $Z_2$.
($S_3$-doublet (singlet) Higgs fields have been assigned
to be odd (even)under the $Z_2$-parity.)
Then we consider the $S_3$-invariant superpotential
\bea
W &=& \mu_D H_I^U H_I^D + \mu_S H_3^U H_3^D+
 m_D N_I N_I + m_S N_3 N_3\nn\\
&+& \lambda_1 H_I^U H_I^D N_3 
+ \lambda_3 H_3^U H_3^D N_3
+ \lambda_2 H_I^U H_J^D (\kappa_{IJ} N_1 + \eta_{IJ} N_2) \nn \\
&+& \lambda_4 (N_3)^3+ \lambda_5 N_I N_I N_3,
\label{SMsinglet}
\eea
where the matrices $\kappa$ and $\eta$ are given by
(\ref{kappa}).
The superpotential (\ref{SMsinglet}) 
has no enhanced continuos symmetry.
Observe that the superpotential (\ref{SMsinglet}) 
is also invariant under
\bea
H_1 &\leftrightarrow& \pm H_2~,~
N_1\to  N_1~,~N_2 \to -N_2.
\eea
Furthermore, the $D$ terms and also the relevant soft supersymmetry
breaking terms have the same discrete symmetry.
This implies that 
$ \langle H_1 \rangle = \langle H_2 \rangle~,~
\langle N_2 \rangle=0$
correspond to a stationary point of the
scalar potential \footnote{The other stationary point
$ \langle H_1 \rangle = -\langle H_2 \rangle$ is physically
equivalent to $ \langle H_1 \rangle = \langle H_2 \rangle$, 
because they can be related 
by a phase rotation of the matter supermutiplets.}.
We, therefore, can expect that
the desired properties
(avoiding the pseudo Goldstones and 
$\langle H_1 \rangle = \langle H_2 \rangle$)
can be satisfied in a wide range of the parameter space.
The masses of the  Higgs multiplets 
depend on the parameters in the scalar potential,
and the presence of multiple Higgs particles
 give rise to FCNCs and CP-violations at the tree-level.
In multi-Higgs models, the tree-level
 FCNCs and CP-violations may cause problems in general.
 However, various dangerous processes in the nonsupersymmetric case
have been  considered in \cite{kubo}, and it has been found that
these processes satisfy the severe experimental upper bounds.
So we may assume that the situation does not change
by supersymmetrization.
A complete analysis on this problem will go beyond
the scope of the present paper, and  we would like to
leave it for future problems. 

The idea to use nonabelian discrete symmetries to
soften the SUSY flavor problem is not new \cite{hall,hamaguchi}.
The only problem is that the symmetries considered in
\cite{hall,hamaguchi}
are explicitly broken at the supersymmetry breaking scale.
Therefore, there is no convincing reason why they should be intact
in the SSB sector. The flavor symmetry recently found in \cite{kubo}
seems to be exact so far: It is  spontaneously broken,
only because the electroweak gauge symmetry
$SU(2)_L\times U(1)_Y$ is spontaneously broken.
It is therefore the most natural to assume that the flavor symmetry
is intact in the SSB sector, too. 
We indeed found in this letter that the flavor
symmetry of \cite{kubo} can soften considerably the SUSY flavor
problem.

\vspace{0.5cm}
\noindent
{\large \bf Acknowledgments}\\
This work is supported by the Grants-in-Aid for Scientific Research 
from the Japan Society for the Promotion of Science (JSPS)
(No. 14540252, 13135210, 14540256, 13640272).

\newcommand{\bi}{\bibitem}

\end{document}